\documentclass[aps,prl,twocolumn,superscriptaddress,letterpaper]{revtex4-1}
\pdfoutput=1
\usepackage{amsmath,amsfonts,amssymb,amsbsy,amscd,amsgen}
\usepackage{float}
\usepackage[usenames,dvipsnames]{xcolor}
\usepackage{graphicx}

\newcommand{\rev}[1]{{#1}}

\renewcommand{\vec}[1]{{\bf #1}}

\newcommand{\butw}{\ensuremath{{\bf u}_{\text{TW}}}}
\newcommand{\bueq}{\ensuremath{{\bf u}_{\text{EQ}}}}
\newcommand{\Rey}{\ensuremath{\mathrm{Re}}}
\newcommand{\Ro}{\ensuremath{\mathrm{Ro}}}

\def\url#1{} % hack to disable urls in bibliography

\begin{document}

\title{The origin of localized snakes-and-ladders solutions of plane Couette
  flow}

\author{Matthew Salewski}
\affiliation{Institut f{\"u}r Mathematik, Technische Universit{\"a}t Berlin, 10623 Berlin Germany}
%\affiliation{Max Planck Institute for Dynamics and Self-Organization, 37077 G{\"o}ttingen, Germany}
\author{John F.  Gibson}
\affiliation{Department of Mathematics and Statistics, University of New Hampshire, Durham, NH 03824, USA}
\author{Tobias M.  Schneider}
\email{tobias.schneider@epfl.ch}
\affiliation{Emergent Complexity in Physical Systems Laboratory (ECPS), {\'E}cole Polytechnique F{\'e}d{\'e}rale de Lausanne, 1015 Lausanne, Switzerland}
%\affiliation{Max Planck Institute for Dynamics and Self-Organization, 37077 G{\"o}ttingen, Germany}

\date{\today}

\begin{abstract}
Spatially localized exact solutions of plane Couette flow are organized in a snakes-and-ladders structure strikingly similar to that observed for simpler pattern-forming partial differential equations. [PRL 104,104501 (2010)]. We demonstrate the mechanism by which these snaking solutions originate from well-known periodic states of the Taylor-Couette system.  They are formed by a localized slug of Wavy-Vortex flow that emerges from a background of Taylor vortices via a modulational sideband instability. This mechanism suggests 
a deep connection between pattern-formation theory and Navier-Stokes flow.
\end{abstract}

\maketitle

The coexistence of laminar and turbulent flows is an issue of long-standing interest, fundamental to the transition process in spatially extended, linearly stable shear flows.
From dynamical systems theory, the discovery of exact invariant solutions of the full nonlinear Navier-Stokes equations has led to much progress in understanding the dynamics of  transitional flows.
Invariant solutions were first studied in the simplified context of small periodic domains. More recently, invariant states with localized support have been computed for spatially extended domains. Examples include localized equilibria and traveling waves in plane Couette flow \cite{Schneider2010a, Schneider2010,Deguchi2013a,Brand2014,Melnikov2014,Gibson2014,Chantry2015,Gibson2016,Reetz2019a}, traveling waves and periodic orbits of plane Poiseuille flow \cite{Gibson2014, Zammert2013, Zammert2014,Mellibovsky2015}, traveling waves in a parallel boundary layer \cite{Kreilos2016, Zammert2016} and periodic orbits of pipe flow \cite{Avila2013,Chantry2014b,Ritter2018}. Such localized solutions intrinsically feature turbulent-laminar coexistence and are thus key to extending the dynamical systems approach to turbulence to the spatiotemporal dynamics of transitional shear flows in extended domains.

The first known localized invariant solutions of plane Couette flow are of special interest because they exhibit homoclinic snaking \cite{Schneider2010,Gibson2016,Knobloch2015}, a characteristic snakes-and-ladders bifurcation structure which relates the localized solutions to more commonly studied periodic solutions. This snakes-and-ladders structure is found in simpler, well-understood pattern forming systems, such as the one-dimensional Swift-Hohenberg equation \cite{Swift1977,Burke2006, Burke2007,Beck2009}. Thus, the localized snaking solutions suggest that the well-developed mathematical analysis of pattern formation systems might carry over to localized solutions in shear flows and thus provide a route toward understanding the mechanisms underlying laminar-turbulent patterns. Despite the striking similarities between plane Couette solutions and solutions of Swift-Hohenberg, the mechanism by which the localized snaking solutions emerge has remained unclear. The origin of the similarity between shear flow solutions and Swift-Hohenberg is thus not fully understood.
In this Rapid Communication we elucidate the origin of the snakes-and-ladders solutions. By rotating the plane Couette system around a spanwise-oriented axis, we demonstrate that localized plane Couette solutions are connected to well-known periodic states of the Taylor-Couette system. They are formed by a localized slug of Wavy-Vortex flow that emerges in a background of Taylor-Vortex flow via a modulational sideband instability.
Since Taylor-Couette is closely related to the Rayleigh-Benard system, the snaking solutions can also be connected to modulated convection rolls \cite{Olvera2017}.

%Approach
In the Swift-Hohenberg model, localized solution branches emerge via a bifurcation from the spatially homogenous background solution which loses linear stability at a critical value of the control parameter \cite{Bergeon2008,Knobloch2015}. The localized states are well understood at small amplitude close to the bifurcation. For plane Couette flow, the laminar state remains linearly stable for all finite Reynolds numbers \cite{Romanov1973}. Consequently, there is no connection of the localized solution branches and the laminar background at which the origin of the localized states could be understood. We thus follow an approach pioneered by Nagata \cite{Nagata1990,Nagata2013} who computed the first spatially periodic invariant solutions of plane Couette flow by exploiting homotopy from Taylor Couette flow. Anti-cyclonic rotation around a spanwise axis destabilizes the laminar flow and allows us to follow the emergence of snaking solutions via a sequence of bifurcations.

\begin{figure}
\includegraphics[width=0.7\columnwidth]{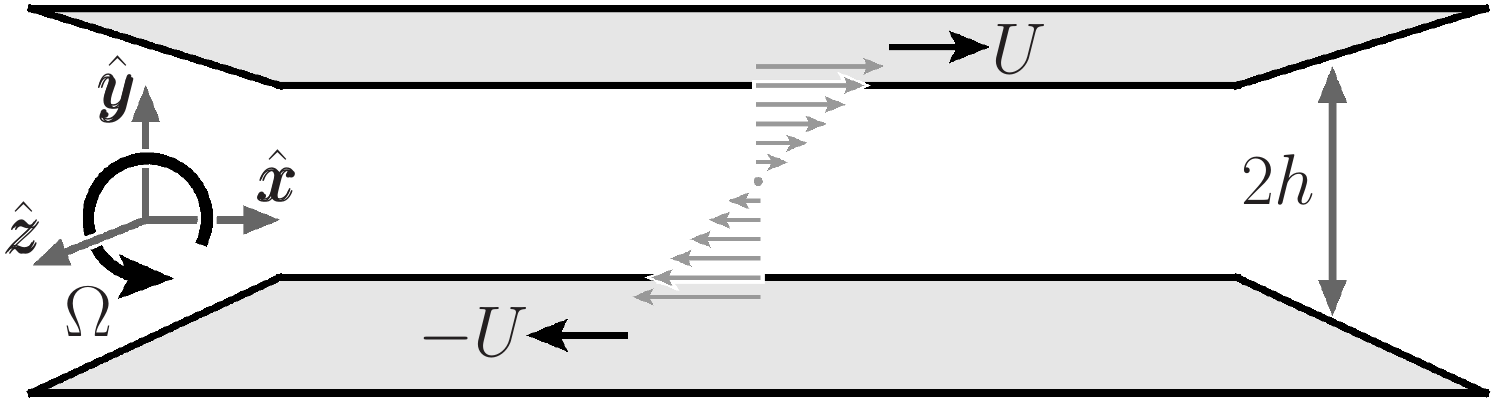}
\caption{(color online) Schematic of plane Couette flow rotating around the spanwise ($\hat{z}$) axis. \label{RPCF_Schematic}}
\vspace{-3mm}
\end{figure}
In rotating plane Couette flow (RPCF) \cite{Mullin2010,Tsukahara2010}, the flow between two parallel walls moving in opposite directions and rotating around a spanwise axis (schematic in Fig.~\ref{RPCF_Schematic}), the velocity field $\vec{u} (\vec{x},t) = [u,v,w] (x,y,z,t)$ evolves under the incompressible Navier-Stokes equations
\begin{equation}
\frac{\partial \vec{u}}{\partial t} + \vec{u} \cdot \nabla \vec{u}	= - \nabla p + \frac {1} {\Rey} \nabla^2 \vec{u} + \Ro (\vec{u} \times \hat{z}),\; \nabla \cdot \vec{u} = 0,\label{NavSt}
\end{equation}
in the domain $V=L_x \times L_y \times L_z$
where $(x, y, z)$ are streamwise, wall-normal and spanwise directions. The boundary conditions are no-slip on the walls $\vec{u}(y=\pm 1) = \pm \hat{x}$ and periodic in $x$ and $z$. 
The two control parameters Reynolds and rotation number are $\Rey = \frac{Uh}{\nu}$ and $\Ro= \frac{2 \Omega h}{U}$, where $U$ is half the relative velocity of the walls, $h$ half the wall separation, $\nu$ the kinematic viscosity and $\Omega$ the rotation frequency around the spanwise axis $\hat{z}$. 
RPCF corresponds to the thin-gap limit of Taylor-Couette flow, the flow between differentially rotated concentric cylinders \cite{Coles1965}. It captures the Coriolis force whose strength is controlled by $\Ro$ but neglects the geometric curvature of the cylinders when the gap width is small relative to the cylinder radius. 
For $0< \Ro <1$, the primary laminar flow loses stability at $\Rey^2 = 107/(\Ro (1-\Ro))$ to streamwise-invariant Taylor-Vortex flow (TVF) \cite{Lezius1976} \footnote{The definition of the Reynolds number used ref.~\cite{Lezius1976} leads to a numerical factor of 2 relative to the Reynolds number used here.}. This secondary state itself becomes unstable to tertiary Wavy-Vortex flow (WVF) exhibiting a streamwise modulation of the flow \cite{Davey1968}.

\begin{figure}[h]
\includegraphics[width=\columnwidth]{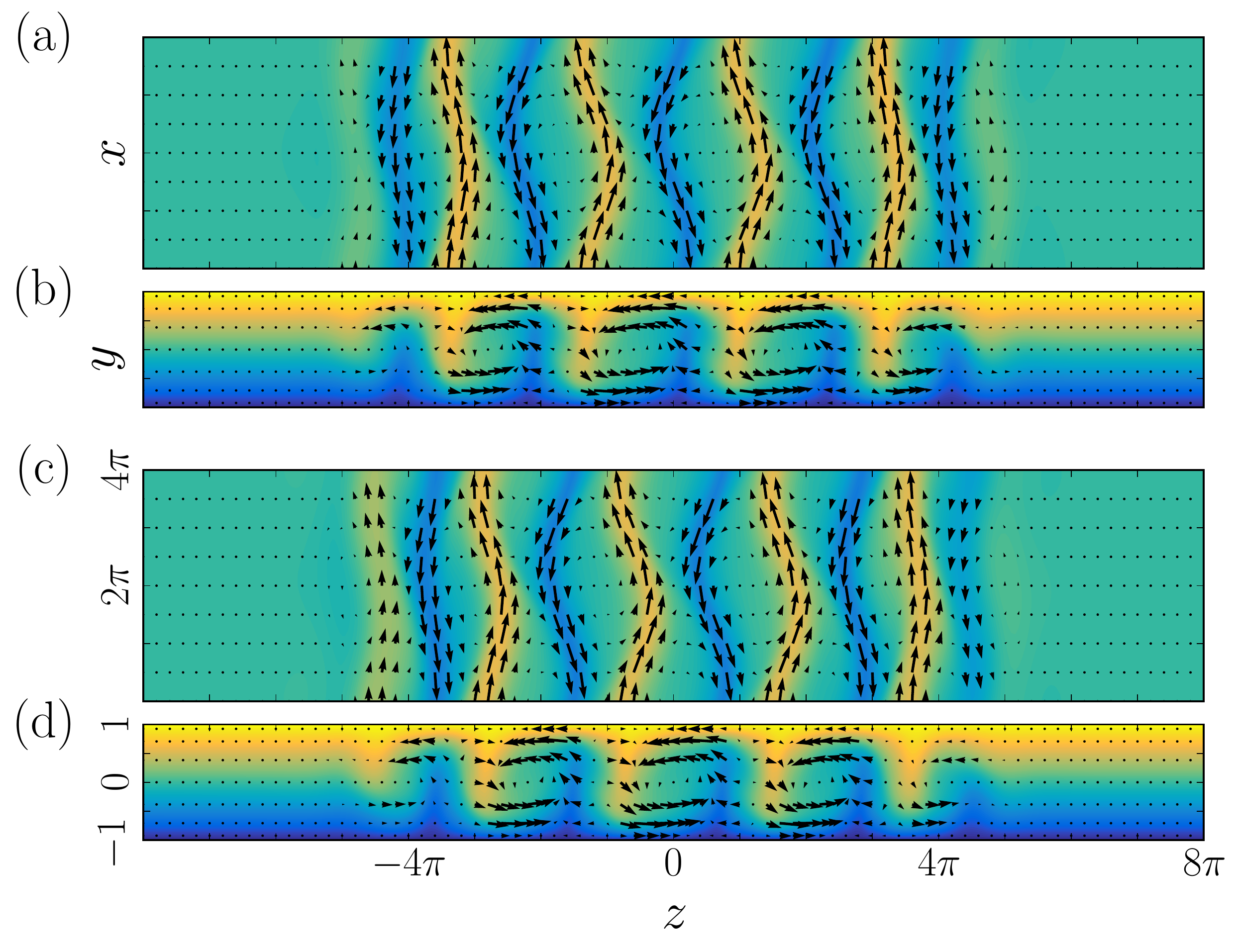}
\caption{(color online) Localized traveling wave $\butw$ (top) and equilibrium $\bueq$ (bottom) followed to finite rotation $\Ro=10^{-4}$. (a),(c) show the velocity in the $y=0$ midplane, with arrows indicating in-plane velocity and the color scale streamwise velocity $u$: dark, medium, light (blue, green, yellow) correspond to $u=[-1,0,1]$. (b),(d) are the streamwise velocity with in-plane velocity indicated by arrows.  \label{SolutionPlots}}
\end{figure}

The snaking solutions were found for the non-rotating case ($\Ro=0$) of eq.~\ref{NavSt} and continued to finite rotation $\Ro$ using a Newton-Krylov-hookstep algorithm combined with quadratic extrapolation in pseudo-arclength \cite{Gibson2008,Gibson2009}. The computations were carried out using the channelflow library \cite{channelflow,Gibson2019} that employs a Fourier-Chebyshev spatial disretization and 3rd-order semi-implicit backward difference time stepping scheme, which we modified to include the Coriolis term explicitly.

Fig.~\ref{SolutionPlots} shows two exact solutions at $\Rey \approx 170$ in a domain of $V = 4\pi\times 2\times 16\pi$ taken from the two snaking 
branches in PCF and continued to anti-cyclonic rotation with $\Ro=10^{-4}$.
The solutions are localized in spanwise direction with the center dominated by wavy roll-streak structures similar to the widely studied periodic 
Nagata, Busse, Clever, Waleffe (NBCW) equilibrium \cite{Nagata1990,Clever1997,Waleffe1998}. The rotation number is small enough that at low $\Rey$ the solutions are hardly modified compared to the non-rotating case, but large enough to destabilize the laminar flow at large but finite $\Rey$. The equilibrium $\bueq$ is stationary, and the traveling wave $\butw$ moves as $[u,v,w](x,y,z,t) = [u,v,w] (x-c_xt, y, z, 0)$ with wave speed $c_x = -0.0016$.  The equilibrium is invariant under inversion $[u,v,w](x,y,z,t) = [-u,-v,-w](-x,-y,-z,t)$ while the traveling wave is shift-reflect symmetric $[u,v,w](x,y,z,t) = [u,v,-w](x+L_x/2,y,-z,t)$. The deviation of the localized solution from laminar flow tapers to magnitude $10^{-3}$ at $z = \pm 8\pi$, showing that for these solutions $L_z = 16\pi$ is an adequate approximation to a spanwise-infinite domain.

\begin{figure*}
\includegraphics[width=\textwidth]{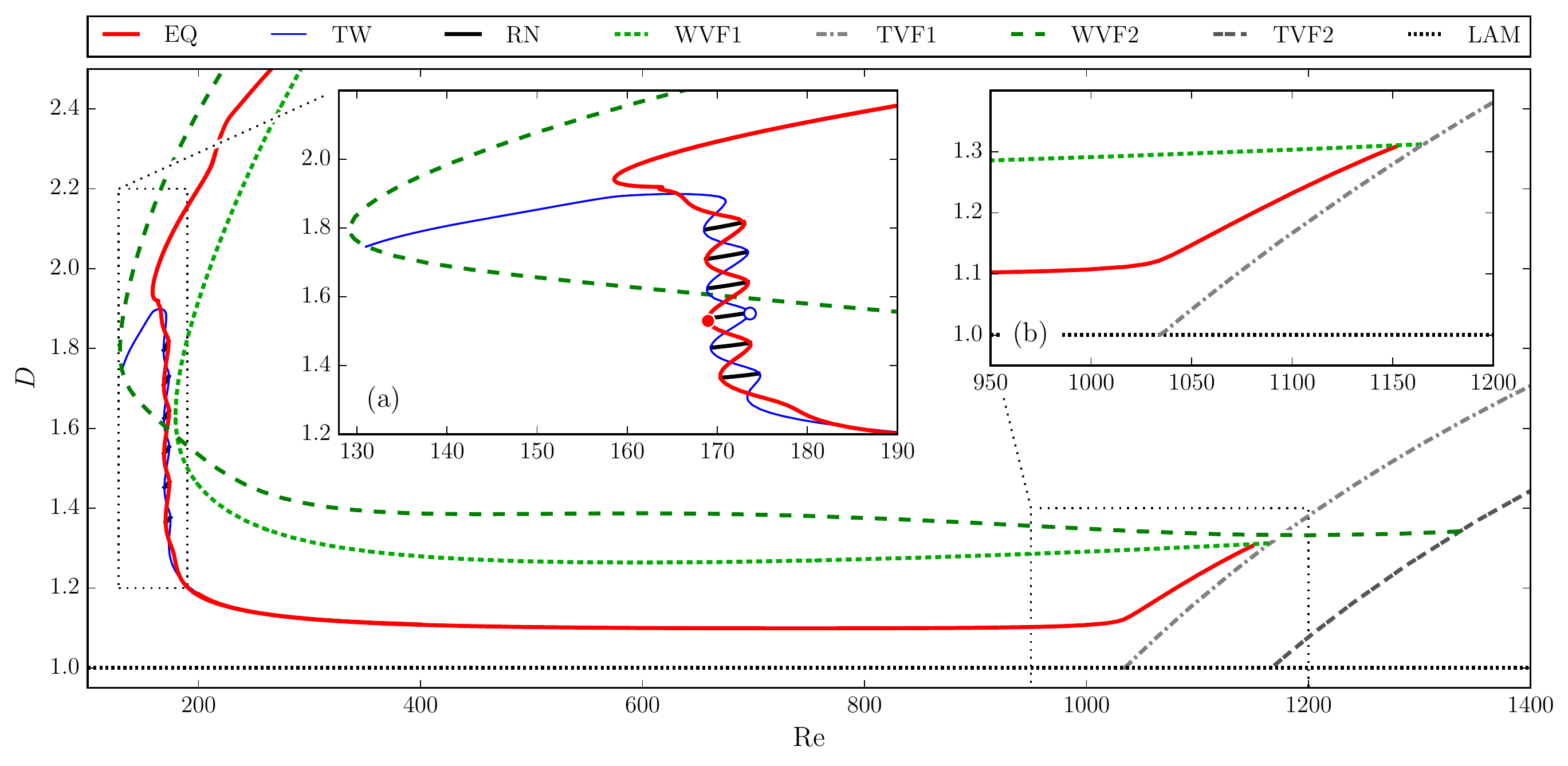}% Images in 100% size
\caption{(color online) Bifurcation diagram of the localized $\butw$, $\bueq$ for fixed $\Ro=10^{-4}$ in the ($\Rey$,$D$) plane with dissipation rate $D=V^{-1} \int|\nabla\times \vec{u}|^2 dx^3$ treated as solution measure. Shown also are the rung states connecting $\butw$ and $\bueq$ as well as spatially periodic TVF and WVF solutions with spanwise wavenumber $\zeta=13$ (TVF1, WVF1) and $\zeta = 8$ (TVF2, WVF2).
Inset (a): The snakes-and-ladders structure observed in the non-rotating system is preserved at finite rotation. The velocity fields for the points marked by the circles are shown in Fig.~\ref{SolutionPlots}. 
Inset (b): Both localized $\butw$ and $\bueq$ emerge via a tertiary bifurcation from the laminar flow: $LAM \rightarrow TVF \rightarrow WVF \rightarrow (\butw, \bueq) $. \label{bifDiag} }
\end{figure*}

Fig.~\ref{bifDiag} shows the parametric continuation of $\bueq$ and $\butw$ in $\Rey$ at fixed rotation $\Ro = 10^{-4}$. On reducing $\Rey$, the branch undergoes homoclinic snaking: in a range of $168<\Rey<175$ a sequence of saddle-node bifurcations leads to spatial growth of the structure by adding one pair of vortices for each wind of the snaking curve \cite{Schneider2010,Gibson2016}. Connecting the entwined snaking branches of $\butw$ and $\bueq$ are additional rung states which have neither the equilibrium nor traveling-wave symmetry and which travel in both the $x$ and $z$ directions. The snakes-and-ladders structure found in non-rotating PCF and strongly similar to Swift-Hohenberg snaking is thus structurally stable. It is not destroyed by rotation; the Coriolis force merely shifts the solutions to slightly smaller $\Rey$ and reduces the width of the snaking curve in $\Rey$. 

For small $\Rey$, rotation does not qualitatively change the bifurcation structure but for higher $\Rey$ a significant modification is observed: instead of remaining separated from the laminar state for all finite $\Rey$ and thereby defying an analysis of their origin as in Swift-Hohenberg, the snaking branches now arise from a tertiary bifurcation from laminar flow. At $\Rey=1035$ the laminar flow is destabilized in a pitchfork bifurcation, breaking the continuous translational symmetry and creating spatially-periodic, streamwise-invariant Taylor Vortex flow (TVF). Being a purely 2D flow, TVF cannot exist without rotation. At $\Rey=1165$ Wavy Vortex Flow (WVF) bifurcates from TVF in a subcritical secondary pitchfork bifurcation giving rise to periodic vortex pairs with wavy modulation in streamwise direction. This secondary state can be followed to $\Ro=0$ and is homotopic to the NBCW equilibrium in PCF as first shown by Nagata.
Close to the onset of WVF the localized branches of $\bueq$ and $\butw$ emerge in a tertiary pitchfork bifurcation that breaks the spanwise periodicity and leads to localization.

\begin{figure}
\includegraphics[width=\columnwidth]{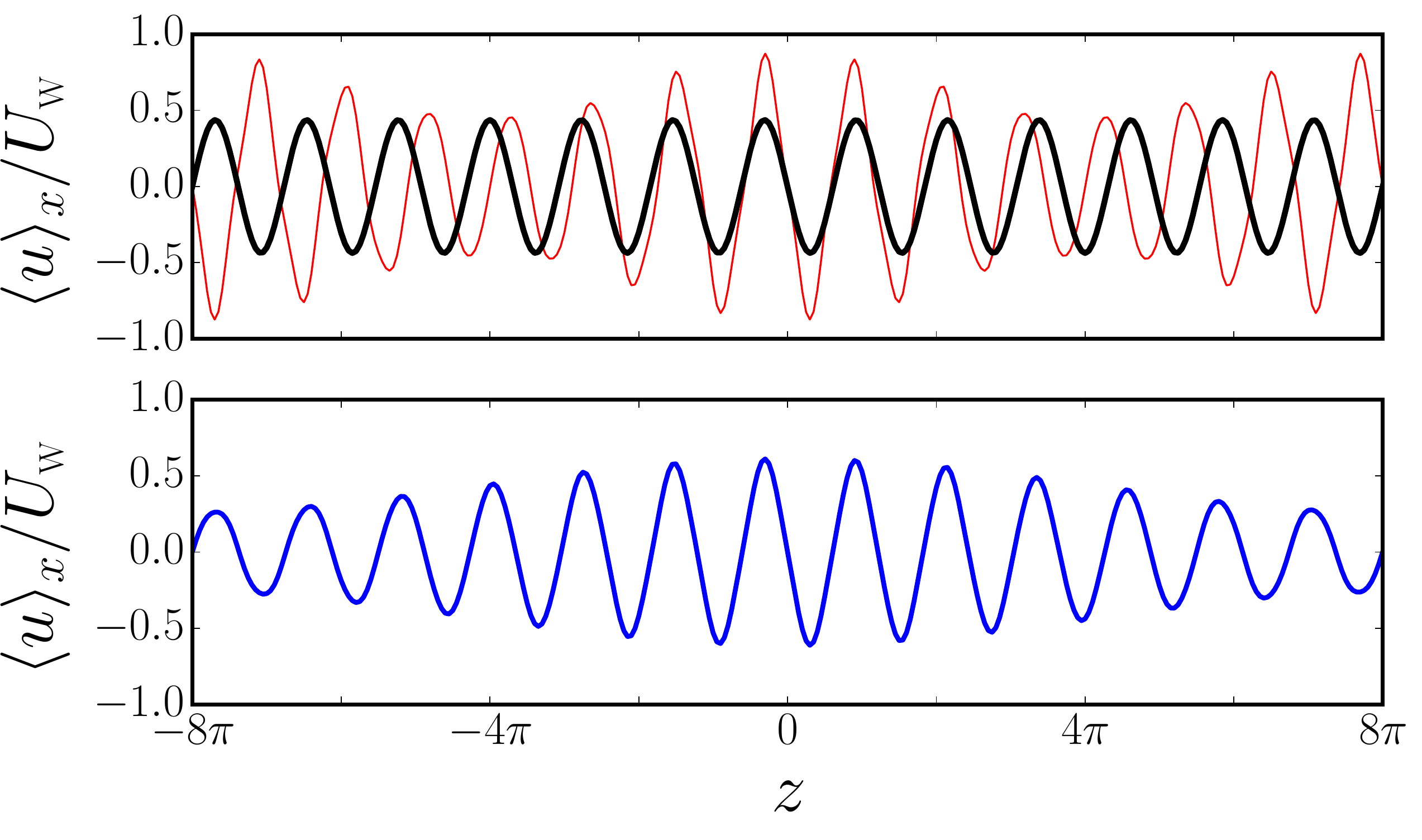}
\caption{(color online) Streamwise-averaged streamwise velocity at the midplane ($y=0$) of critical WVF (top, thick), and the neutral eigenmode (top, thin) of the pitchfork bifurcation creating the localized $\bueq$ branch. The detuning of the spatial frequency of the eigenmode relative to the base causes phase-interference and thereby an amplitude modulation leading to localization (bottom). The eigenmode giving rise to $\butw$ has an almost identical structure but instead preserving of inversion symmetry it is shift-and-reflect symmetric.
\label{Sideband}}
\end{figure}

To analyze the mechanism by which the bifurcation creates localization, we compute eigenvalues and eigenmodes using Arnoldi iteration. The localized branches emerge in a pitchfork bifurcation resulting from two degenerate real eigenvalues crossing the imaginary axis. Of the two neutral eigenmodes, one gives rise to the localized equilibrium and the other the localized traveling wave. Fig.~\ref{Sideband} shows the mechanism by which discrete translational symmetry is broken and localization emerges, in terms of the streamwise-averaged streamwise velocities in the $y=0$ midplane for both the periodic state and the neutral eigenmode. While the periodic WVF (thick line) has a spatial frequency of $\zeta=13$ in units of $1/L_z$, i.e. 13 periodic vortex pairs, the eigenmode (thin line) is dominated by Fourier modes with frequency $\zeta=13 \pm 1 = 12,14$ and contains 14 vortex pairs. Adding a small amount of the spatially-detuned eigenmode to the WVF base state leads to phase interference: the amplitude increases where the eigenmode is in phase with the base state and reduces when it is out of phase. As a result a beating pattern or amplitude modulation with periodicity of the computational domain emerges and leads to localization. This phenomenology is characteristic of a modulational sideband instability of Eckhaus type \cite{Tuckerman1990}. 
As the spanwise length of the computation domain increases, the detuning of the eigenmodes relative to the base pattern decreases. As a result, the bifurcation point of the localized branches approaches that of the WVF, and in the limit of a spanwise-infinite domain, the WVF and the localized branches bifurcate together from TVF \cite{Bergeon2008}. 

\begin{figure}
\includegraphics[width=\columnwidth]{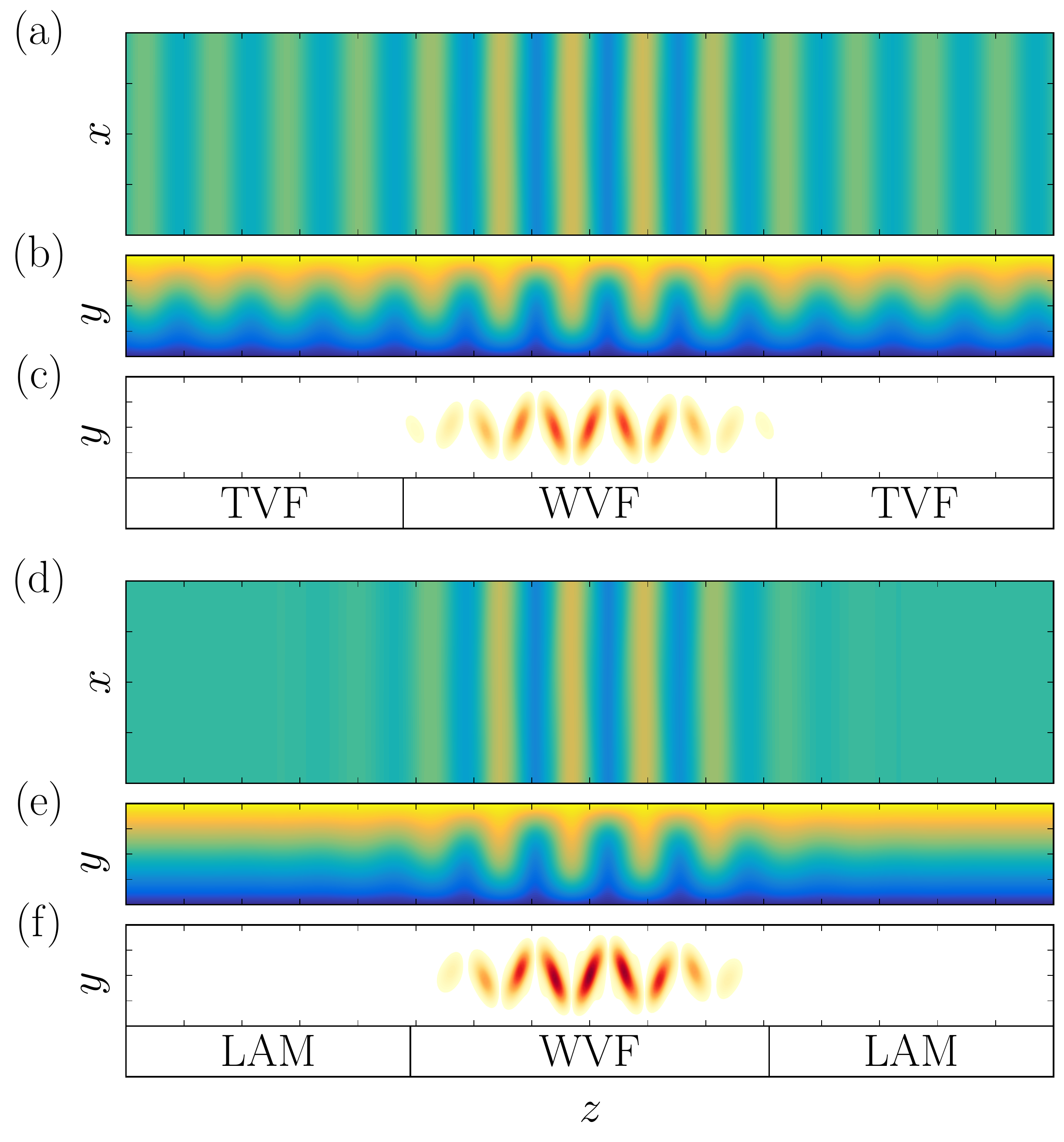}
\caption{(color online) The localized solution before and after the TVF-onset, at $\Rey = 1050$ (a--c) and $\Rey = 1000$ (d--f), respectively. (a,b,d,e): visualization as in Fig.~\ref{SolutionPlots}. (c,f): filtered spanwise energy density $E(y,z)=\langle\vec{u}^2_{3D}\rangle_{x}/2$ in the range $E=[0\,(\text{light}),10^{-3}(\text{dark})]$. The core of the localized branch resembles streamwise-modulated 3D WVF while the background is formed by 2D TVF. On reducing $\Rey$, the TVF background vanishes, leaving the localized core embedded in a laminar background. \label{background_compare}}
\end{figure}

The modulated pattern shown in Fig.~\ref{Sideband} (bottom) transitions to
a strongly localized pattern under continuation to lower Reynolds
numbers. Fig.~\ref{bifDiag}(b) details the region of this transition,
beginning with the modulating bifurcation off WVF at $\Rey = 1150$ and
strong localization developing by $\Rey \approx 1000$. 
Fig.~\ref{background_compare} (a,b,c) shows $\bueq$ at $\Rey = 1050$. 
The solution here appears to be formed from central core of 3D 
streamwise-modulated WVF surrounded by weaker 2D streamwise-invariant 
TVF. To highlight the distinction between these regions, 
fig.~\ref{background_compare}(c) shows the energy of the 3D streamwise 
modulation, $E(y,z) = \langle \vec{u}^2_{3D}\rangle_x/2$, where 
$\vec{u}_{3D} = \vec{u} - \langle \vec{u} \rangle_x$ and angle brackets 
indicate averaging. The energy of the 3D streamwise modulation
is clearly confined to the central region. Consequently, the state 
formed by the sideband instability can be interpreted as a localized 
slug of WVF embedded in a TVF background. On reducing $\Rey$, the TVF 
background reduces in amplitude and eventually vanishes at the 
$\Rey =  1035$ TVF bifurcation point. Fig.~\ref{background_compare} (d,e,f)
show $\bueq$ just past this point, at $\Rey=1000$. The localized WVF core, 
now embedded in a laminar background, survives and forms the localized 
state which then proceeds to homoclinic snaking.

\begin{figure}
\includegraphics[width=\columnwidth]{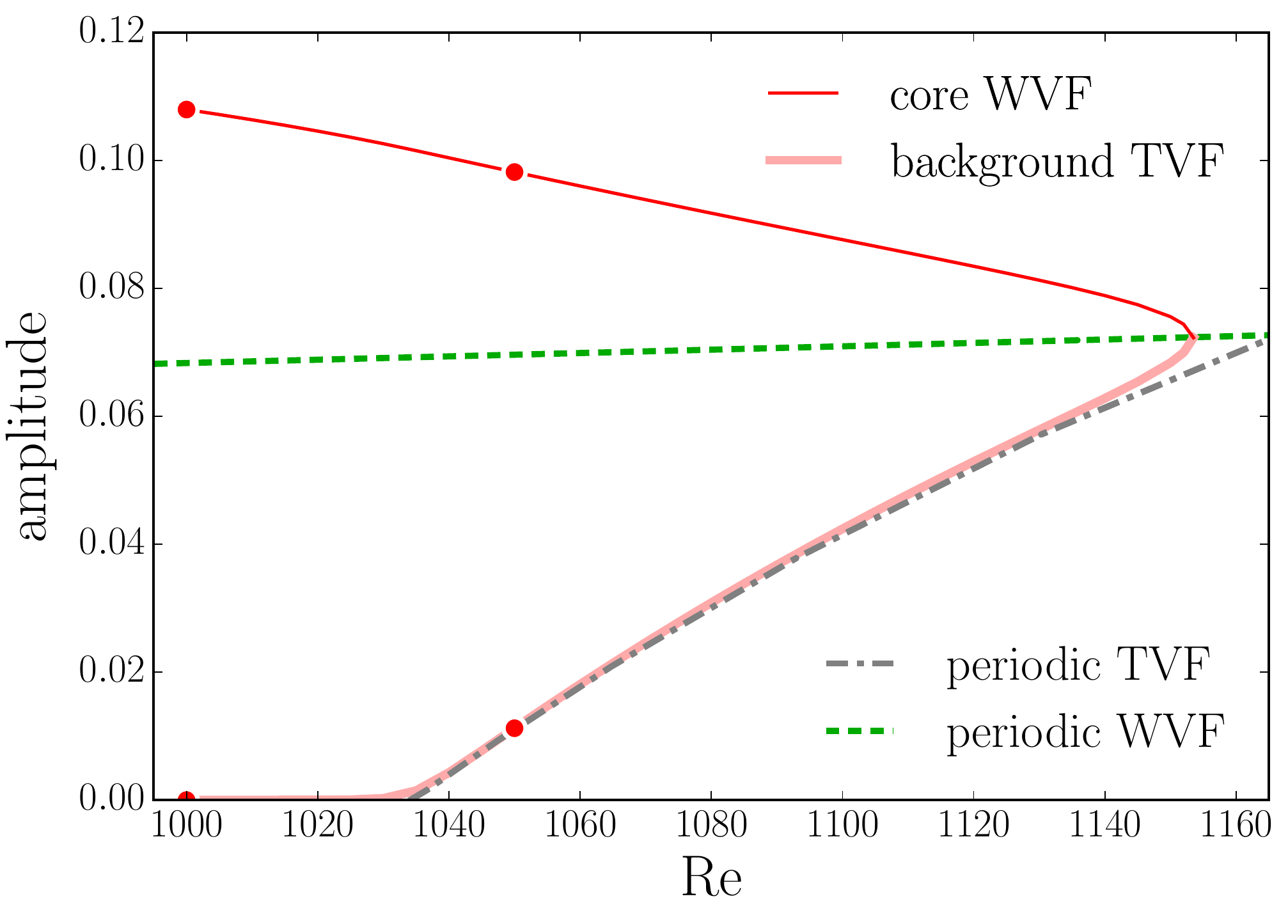}
\caption{Amplitudes of the 3D WVF core, the 2D TVF background,
and the periodic WVF and TVF solutions as $\Rey$ is reduced from the 
localizing bifurcation at $\Rey=1150$. Amplitudes are quantified by 
the spanwise maxima of the $xy$-average energy density 
$\langle u^2\rangle_{xy} (z)/2$ near $z=0$ (WVF core) and $z=\pm L_z/2$ (TVF background), and across the whole spanwise domain for the periodic solutions. The amplitude of the TVF background closely follows TVF and vanishes at $\Rey < 1035$. The amplitude of the core however decouples from WVF and increases as $\Rey$ is reduced away from the bifurcation point. The solid circles denote the solutions shown in Fig.~\ref{background_compare}\label{amplitudeModulation}}
\end{figure}

Fig.~\ref{amplitudeModulation} shows how the amplitude of the WVF core and 
the TVF background change as $\Rey$ is reduced away from the bifurcation point.
%The amplitudes are quantified by the maxima of the $xy$-average energy density
%$\langle \vec{u}^2\rangle_{xy} (z)$, near $z=0$ (WVF core) and $z=\pm L_z/2$ (TVF background). 
The background TVF amplitude follows the amplitude of the periodic TVF solution until it vanishes at the primary instability at $\Rey<1035$, whereas the localized WVF core strengthens in amplitude and is largely unaffected by the changing background.

Localized states in driven dissipative systems typically resemble a slug of patterned state embedded in a uniform translationally-invariant background. 
% Adding comments of the referee
\rev{
Examples in shear flows include localized states in a laminar background emerging in long-wavelength modulational instabilities that create a laminar 'hole' in a spatially periodic state. Those have been identified in plane Couette \cite{Melnikov2014}, plane Poiseuille \cite{Mellibovsky2015}, pipe \cite{Chantry2014b} and magnetized Taylor-Couette flow \cite{Guseva2015}.
}
For the subcritical Swift-Hohenberg equation such states \rev{with uniform background} can be described in terms of spatial dynamics where the time-independent version of the 1D PDE is treated as a dynamical system in space \cite{Knobloch2015}. The uniform state is a fixed point while the periodic patterned state is represented by a periodic orbit. The localized state can be understood as a homoclinic orbit of the fixed point that visits the neighborhood of the periodic orbit before returning to the fixed point.

The localized snaking solutions in RPCF however do not emerge from a spanwise uniform state but from the spanwise periodic TVF. Consequently, the suggested spatial dynamics picture is that of a homoclinic orbit to the periodic orbit corresponding to TVF, which passes in the neighborhood of the orbit corresponding to WVF. Once $\Rey$ is reduced, the TVF orbit shrinks and collides with the fixed point of laminar flow. Now the scenario of a homoclinic orbit connecting a fixed point and one periodic orbit is recovered. \rev{An analogous scenario where a localzed slug of one patterned state emerges in a background of a second patterned state does not occur in the commonly studied Swift-Hohenberg equation with cubic-quintic nonlinearity but has recently been observed when a cubic-quintic-septic nonlinearity is considered \cite{Knobloch2019}.}

On a finite domain the snaking may terminate by reconnecting to a periodic state when the localized solution has grown to the size of the domain. For the parameters chosen here, the equilibrium branch $\bueq$ terminates on a periodic WVF state, corresponding to the NBCW equilibrium at \Ro=0. Note however, that it reconnects to a WVF state with 8 vortex pairs while it emerges from a WVF with 13 vortex pairs, a spanwise wavelength close to the critical TVF wavelength \cite{Lezius1976}. On a periodic domain, WVF of different periodicity compatible with the periodic boundary conditions may thus be connected smoothly.

%Summary
We have elucidated the origin of spatially localized exact invariant solutions that exhibit homoclinic snaking and thereby suggest a deep connection between localized states in shear flows and well-studied pattern forming mechanisms. By adding anti-cyclonic rotation, the snakes-and-ladders solutions in plane Couette flow are found to be formed by localized slugs of Wavy Vortex flow that emerges in a background of Taylor Vortex flow via a modulational sideband instability. Unlike the TVF background they emerge from, these slugs of WVF survive for vanishing rotation and form the localized snaking solutions in plane Couette flow.

%\begin{acknowledgments}
%\end{acknowledgments}

\vspace{-3mm}
%\bibliography{./library}
%

\end{document}